\documentstyle[prb,aps,multicol,psfig]{revtex}
\begin{document}
\draft
\tighten

\title{Kinetics of spin coherence of electrons in an undoped semiconductor
quantum well}
\author{M. W. Wu$^*$ and H. Metiu}
\address{Chemistry Department, University of California, Santa Barbara,
CA 93106}

\date{Scheduled to publish in PRB Jan. 15, 2000 issue}
\maketitle
\begin{abstract}
We study the kinetics of spin coherence of optically excited
electrons in an undoped insulating ZnSe/Zn$_{1-x}$Cd$_x$Se quantum well
under moderate magnetic fields
in the Voigt configuration. After clarifying the optical coherence and the
spin coherence, we build the kinetic Bloch equations and
calculate dephasing and relaxation kinetics
of laser pulse excited plasma due to statically
screened Coulomb scattering and electron hole spin exchange.
We find that the Coulomb scattering can not cause the spin dephasing,
and that the electron-hole spin exchange is the main mechanism of the spin
decoherence. Moreover the beat frequency in the Faraday rotation
angle is determined mainly by the Zeeman splitting, red shifted by the
Coulomb scattering and the electron hole spin exchange.
Our numerical results are in agreement
with experiment findings. A possible scenario for the contribution
of electron-hole spin exchange to the spin dephasing of the $n$-doped
material is also proposed.
\end{abstract}

\begin{multicols}{2}
\narrowtext

\section{Introduction}
Studies of ultrafast nonlinear optical spectroscopy in semiconductors
have attracted numerous interest both experimentally and theoretically
during the past 20 years.\cite{proce,shah,ufpxi} Most of these studies
are focused on the optical coherence and the studies of spin coherence are
relatively rare. Recently, ultrafast nonlinear optical
experiments\cite{dam,wagner,baum,herb,buss1,crook,buss2,kikk1,kikk2,kikk3}
have shown
that the spin coherence which is optically excited by laser pulse can last
much longer than optical coherence. For undoped ZnSe/ZnCdSe quantum wells,
it is found in the experiment that the spin coherence can last up to
15-20\ ps\cite{kikk1} where as for undoped bulk GaAs, it lasts about
600\ ps.\cite{kikk3} For $n$-doped material, the spin coherence can last
up to three orders of magnitude longer than the undoped sample which makes
8\ ns for ZnSe/ZnCdSe quantum well\cite{kikk1} and 100\ ns for bulk
GaAs.\cite{kikk3} These discoveries have given rise
to an emerging interest within the physics
and electronic engineering communities in using electronic spins for the
storage of coherence and also have stimulated the optimism that the coherent
electrons will finally be realized as a basis for quantum computation.

The electron spin coherence can be directly observed by
femtosecond time-resolved
Faraday rotation (FR) in the Voigt configuration. In that configuration
a moderate magnetic field is applied normal to the growth axis of the
sample. The coherence is introduced by a circularly polarized pump pulse
that creates electrons and holes with an initial spin orientation normal to
the magnetic field. Then the magnetic field causes the electron spin to
flip back and forth along the growth axis which makes the net spin precess
about the magnetic field. The hole spin is kept along the
growth axis direction of the quantum well as will be discussed below.
After a certain delay time $\tau$, a linearly polarized probe pulse
is sent into the sample along a slightly different direction from
the pump pulse and by measuring the
FR angle, one can sensitively detect the net spin of electrons associated
with the delay time $\tau$. This method has proven to be extremely
successful in measuring the coherent spin evolution and spin
dephasing.\cite{dam,wagner,kikk1,kikk2,kikk3}

In order to further extend the spin coherence time, it is important
to understand the physics of spin dephasing. While there is extensive
theoretical study and understanding of the optical dephasing,\cite{haug}
the theoretical investigation on the spin dephasing is relatively limited,
nevertheless of much longer history.\cite{meier}
The early work includes that of Elliott in 1954,
who discussed the spin relaxation\cite{elliott} induced by
lattice and impurity scatterings by taking into account
the spin-obit effects.\cite{yafet}
Later, in 1975 Bir {\em et al.} calculated the spin dephasing
using a model Hamiltonian describing the electron-hole spin exchange (EHSE)
by considering Coulomb scattering between electron and hole, combined
with the spin-obit-coupling induced band mixing.\cite{bir}
In that paper by using Fermi Golden rule,
Bir {\em et al.} proposed a spin
relaxation rate  which is proportional to the hole density. It was not
until 90's that experimentalists found such an effect and claim the
EHSE is important in the intrinsic and
$p$-doped semiconductors.\cite{dam,wagner} For $n$-doped samples, however,
as the density of the electrons is much higher than that of the holes,
the holes recombine with electrons in a time
much shorter than the measured spin dephasing time. As the predicted spin
dephasing rate induced by EHSE is proportional to hole density,\cite{bir}
it is therefore suggested that for $n$-doped sample, the dephasing
mechanism is unclear.\cite{kikk1}
Recently Linder and Sham\cite{linder} presented a theory of
the spin coherence of excitons by studying Bloch equations.
However, they did not discuss the dephasing mechanisms explicitly, and
instead described all the dephasing by using phenomenological relaxation
times.

In this paper, we present a model to study
the kinetics of  spin precession of a femtosecond laser-pulse excited dense
plasma in a quantum well in the framework of the semiconductor Bloch
equations combined with carrier-carrier scattering in the Markovian
limit. Non-Markovian effects are not important in our case, as the
width of the laser pulse is large (100\ fs) and the time scale of spin
dephasing is very long. The main purpose of this paper is
to understand the spin decoherence. It has been well known
that both carrier-carrier
Coulomb scattering and carrier-phonon scattering play significant roles
in the optical dephasing. For the 2D carrier density around 10$^{11}$\ cm$^
{-2}$ in the experiment,\cite{kikk1} the main carriers are electron-hole
plasma and the Coulomb scattering gives a fast dephasing of the optical
coherence with dephasing times ranging from tens of femtoseconds to
subpicoseconds,\cite{shah} depending on the strength and
width of the pump pulse and/or the density of
doping, which affects the building up of the screening.\cite{haug,vu}
Besides the fast dephasing due to Coulomb scattering, carrier-phonon
scattering also contributes to the optical dephasing with the dephasing time
being around ten picoseconds and carrier-density independent.
For spin coherence, the experimental evidence
that the spin dephasing depends strongly on the carrier
density\cite{kikk1,kikk3}
clearly rules out the possibility that the main mechanism of the spin
dephasing is due to carrier-phonon scattering. Therefore we focus
on the effect of carrier-carrier scattering.
We distinguish the spin coherence from the optical coherence and study the
roles of Coulomb scattering and EHSE to the spin dephasing
separately. We find that the pure Coulomb scattering---although
it destroys the optical coherence
strongly---does not contribute to the spin dephasing at all, and that EHSE
is the main mechanism of the spin decoherence.

Our paper is organized as follows:  we present our model and
kinetic equations in Sec.\ II. Then in Sec.\ III we present
the numerical results for an undoped ZnSe/Zn$_{1-x}$Cd$_x$Se quantum well.
A conclusion of our main results and a discussion of the theory for the
$n$-doped sample are given in Sec. IV.

\section{Model and kinetic equations}
\subsection{Model and Hamiltonian}

We start our investigation of a quantum well with its growth axis in
the $z$-direction. A moderate magnetic field $B$ is applied in the
$x$-direction. Landau quantization is unimportant for the magnetic field
$B$ in our investigation. We consider the conduction band (CB) and
the heavy hole (hh) valence band (VB). Due to the presence of the magnetic
field, the spins of electrons and holes are no longer
degenerate and therefore each band is
further separated into two spin bands with spin $\pm 1/2$ for electrons in
the CB and $\pm 3/2$ for those in the hh VB. It is noted that these spin
eigenstates are defined with respect to the $z$-direction.

In the presence of the moderate magnetic field and with the interactions
with a coherent classical light field, the Hamiltonian for electrons
in the CB and VB is given by
\begin{eqnarray}
H&=&\sum_{\mu k\sigma}\varepsilon_{\mu k}c^\dagger_{\mu k\sigma}c_{\mu k\sigma}
+g\mu_B{\bf B}\cdot\sum_{\stackrel{\mu k}{\sigma\sigma^\prime}}
{\bf S}_{\mu \sigma\sigma^\prime}
c^\dagger_{\mu k\sigma}c_{\mu k\sigma^\prime}\nonumber\\
\label{hal}
&&\mbox{}+H_E+H_I\ ,
\end{eqnarray}
with $\mu=c$ and $v$ standing for the CB and the VB respectively.
$\varepsilon_{vk}$ is the energy spectrum of
an electron in the VB (CB) with momentum $k$. It is noted that $k$ here
stands for a momentum vector in the $x$-$y$ plane. $\varepsilon_{vk}=-E_g/2
-k^2/2m_h\equiv -E_g/2-\varepsilon_{hk}$
and $\varepsilon_{ck}=E_g/2+k^2/2m_e\equiv E_g/2+\varepsilon_{ek}$
with $m_h$ and $m_e$ denoting effective
masses of hh and electron separately. $E_g$ is unrenormalized
band gap and $\sigma$ is the spin index. For
electron in the CB, $\sigma=\pm 1/2$ and for electron in the hh VB,
$\sigma=\pm 3/2$. $\mu_B$ is Bohr magneton.
 ${\bf S}_\mu$ are the spin matrices with ${\bf S}_c$ being
spin $1/2$ matrices for electrons and ${\bf S}_v$ being spin 3/2 matrices
for holes. 

$H_E$ in Eq.\ (\ref{hal}) denotes the dipole coupling with the light field
$E_\sigma (t)$ with $\sigma=\pm$ representing the circular polarized
light. Due to the selection rule the electrons in the spin $3/2$ ($-3/2$)
hh band can only absorb a left (right) circular polarized photon and go
to the spin $1/2$ ($-1/2$) CB. Therefore
\begin{eqnarray}
H_E&=&-d\sum_{k} [E_-(t)c^\dagger_{ck\frac{1}{2}}c_{vk\frac{3}{2}}+H.c.]
\nonumber\\
&&\mbox{}-d\sum_k [E_+(t)c^\dagger_{ck-\frac{1}{2}}
c_{vk-\frac{3}{2}}+H.c.]\ .
\end{eqnarray}
In this equation $d$ denotes the optical-dipole matrix element. The light
field is further split into $E_\sigma(t)=E^0_\sigma(t)\cos(\omega t)$ with
$\omega$ being the central frequency of the coherent light pulse. $E_\sigma
^0(t)$ describes a Gaussian pulse $E_\sigma^0 e^{-t^2/\delta t^2}$ with
$\delta t$ denoting the pulse width.

$H_I$ is the interaction Hamiltonian. As said before, we focus on the
carrier-carrier scattering. Therefore $H_I$ is composed of Coulomb scattering
and EHSE\cite{ehse} with the later being much weaker than
the Coulomb scattering.\cite{single}
\begin{eqnarray}
H_I&=&\frac{1}{2}\sum_{\stackrel{\stackrel{\mu\nu}{kk^\prime q}}
{\sigma\sigma^\prime}}V_q c^\dagger_{\mu k+q\sigma}c^\dagger_{\nu k^\prime
-q\sigma^\prime}c_{\nu k^\prime\sigma^\prime}c_{\mu k\sigma}\nonumber\\
&&\mbox{}+\frac{1}{2}\sum_{\stackrel{\stackrel{\mu\not=\nu}{kk^\prime q}}
{\sigma\sigma^\prime}}U_q \sigma\sigma^\prime
c^\dagger_{\mu k+q\sigma}c^\dagger_{\nu k^\prime
-q\sigma^\prime}c_{\nu k^\prime\sigma^\prime}c_{\mu k\sigma}\nonumber\\
\label{hee}
&&\mbox{}+\frac{1}{2}\sum_{\stackrel{\stackrel{\mu\not=\nu}{kk^\prime q}}
{\sigma\sigma^\prime}}U^\prime_q\sigma\sigma^\prime c^\dagger_{\mu k+q\sigma}
c^\dagger_{\nu k^\prime-q\sigma^\prime}c_{\mu k^\prime\sigma}c_{\nu k\sigma
^\prime}\ .
\end{eqnarray}
The first term of Eq.\ (\ref{hee}) is the ordinary Coulomb interaction.
The second term describes the ``direct'' EHSE which scatters an
electron in the band $\mu$
with spin $\sigma$ and momentum $k$ to the same band
with spin $\sigma$ and momentum $k+q$ and in the
mean time which scatters an electron from the different band $\nu$ ($\not=
\mu$) with spin $\sigma^\prime$ and momentum $k^\prime$ back to that
band with spin $\sigma^\prime$ and momentum $k^\prime-q$.
The last term is ``exchange'' EHSE which scatters an electron
in the band $\mu$
with spin $\sigma$ and momentum $k^\prime$ to the different band $\nu$
($\not=\mu$) with spin $\sigma^\prime$ and momentum $k^\prime-q$ and in the
mean time which scatters an electron from the band $\nu$ with spin
$\sigma^\prime$ and momentum $k$ back to band $\mu$ with spin $\sigma$ and
momentum $k+q$.
It is noted here that $\sigma$ and $\sigma^\prime$ stand
for $\pm 1/2$ when they are at the CB and $\pm 3/2$ when they are at the VB.
As the exchange effect involves a form factor which
consists the overlap of the wavefunctions of the CB and VB, $U_q^\prime$
is much smaller than $U_q$. Moreover, it will be shown later that
the exchange effect is also energetically unfavorable as it scatters
carriers across the band. Therefore the direct EHSE is the dominant effect.

For Voigt configuration, as ${\bf B}$ is along the
$x$ direction and $(S^x_v)_{\pm \frac{3}
{2},\pm\frac{3}{2}}\equiv 0$, one can see directly from
Eq.\ (\ref{hal}) that the spin of hh cannot
be flipped by the magnetic field. We point here that this
is only true when the magnetic field is small compared to the hh-light hole
splitting and hence the flip between the hh and the light hole can be
neglected in the timescale of the discussion.
As ${\bf S}_c^x=\sigma^x/2$ with $\sigma^x$ standing for Pauli
matrix, it is therefore straightforward to see from the Hamiltonian that
the magnetic field causes the CB electron spin to flip and flop.

It can be seen directly from the Hamiltonian $H_E$ that the laser pulse
introduces the optical coherences into the system which are built
between the CB and the VB with same spin
direction: $p_{k\frac{1}{2}\frac{3}{2}}\equiv\langle
c^\dagger_{vk\frac{3}{2}}c_{ck\frac{1}{2}}\rangle$ and $p_{k-\frac{1}{2}
-\frac{3}{2}}\equiv\langle c^\dagger_{vk-\frac{3}{2}}
c_{ck-\frac{1}{2}}\rangle$. In the mean time, due to the presence
of the magnetic field in Eq.\ (\ref{hal}), these optical coherences
may further transfer coherence to
$p_{k-\frac{1}{2}\frac{3}{2}}\equiv\langle
c^\dagger_{vk\frac{3}{2}}c_{ck-\frac{1}{2}}\rangle$, $p_{k\frac{1}{2}
-\frac{3}{2}}\equiv\langle c^\dagger_{vk-\frac{3}{2}}
c_{ck\frac{1}{2}}\rangle$ and $\rho_{cck\frac{1}{2}-\frac{1}{2}}
\equiv\langle c^\dagger_{ck-\frac{1}{2}}c_{ck\frac{1}{2}}\rangle$, with
the first two being the coherence between the CB and VB with opposite
spin directions and the last one being the coherence between two CB's with
opposite spin directions. While it is well known that optical coherence
is represented by $p_{k\frac{1}{2}\frac{3}{2}}$ and
$p_{k-\frac{1}{2}-\frac{3}{2}}$, we will show later that the spin
coherence of electrons is represented by $\rho_{cc
k\frac{1}{2}-\frac{1}{2}}$. When there is no dephasing effect added on
this term, the electron spin precession will last forever. Moreover, when
$\rho_{cc k\frac{1}{2}-\frac{1}{2}}$ decays to zero, there is no electron
spin precession.
Therefore we refer it as spin coherence in the following.
$P_{k-\frac{1}{2}\frac{3}{2}}$
and $P_{k\frac{1}{2}-\frac{3}{2}}$, which describe the coherence between
the states with the optical transition between them being forbidden by the
selection rule, may play certain role in the optical dephasing and we
will refer them hereafter as forbidden optical coherences.
Finally there is also a coherence between two
VB's with opposite spin directions: $\rho_{vvk\frac{3}{2}-\frac{3}{2}}
\equiv\langle c^\dagger_{vk-\frac{3}{2}}c_{vk\frac{3}{2}}\rangle$. In the
absence of spin flip of the hh, this coherence is much weaker than
the other coherences as it then can only be excited by the laser pulse
through the coupling to the forbidden coherence. Therefore in the present
discussion of undoped material, we do not include it in our model.

\subsection{Kinetic equations}

We build the semiconductor Bloch equations for the quantum well by
nonequilibrium Green function method\cite{haug} as follows:
\begin{equation}
\label{kin}
\dot\rho_{\mu\nu,k,\sigma\sigma^\prime}=\dot\rho_{\mu\nu,k,\sigma
\sigma^\prime}|_{\mbox{coh}}+\dot\rho_{\mu\nu,k,\sigma
\sigma^\prime}|_{\mbox{scatt}}\ .
\end{equation}
Here $\rho_{\mu\nu,k,\sigma\sigma^\prime}$ represents a single particle
density matrix with $\mu$ and $\nu=c$ or $v$. The diagonal elements
describe the carrier distribution
functions $\rho_{\mu\mu,k,\sigma\sigma}=f_{\mu k\sigma}$ of the band $\mu$,
the wave vector $k$ and the spin $\sigma$. It is further noted that
$f_{ck\sigma}\equiv f_{ek\sigma}$ represents the electron
distribution function with
$\sigma=\pm \frac{1}{2}$ and $f_{vk\sigma}=1-f_{hk\sigma}$
with $f_{hk\sigma}$ denoting the hh distribution function and $\sigma=\pm
\frac{3}{2}$. The off-diagonal elements describe the
inter spin-band polarization components (coherences) we defined
at the end of the previous subsection with $\rho_{cv,k,\sigma\sigma^\prime}
=p_{k\sigma\sigma^\prime}=P_{k\sigma\sigma^\prime}e^{-i\omega t}$
for the inter CB-VB polarization and
$\rho_{cc,k,\frac{1}{2}-\frac{1}{2}}$ for the spin coherence. It is
noticed here that for $P_{k\sigma\sigma^\prime}$, the first spin
index $\sigma$ always corresponds to
the spin index of the electron in the CB
($\pm 1/2$) and the second spin index
$\sigma^\prime$ always corresponds to that of the hh VB ($\pm 3/2$).

The coherent part of the equation of motion for the electron distribution
function in the rotating wave approximation is given by
\begin{eqnarray}
&&\left.\frac{\partial f_{ek\sigma}}{\partial t}\right|_{\mbox{coh}}
\nonumber\\
&&=d\delta_{\sigma\frac{1}{2}}\mbox{Im}[E^{0\ast}_{-}(t)P_{k\sigma
\frac{3}{2}}]+d\delta_{\sigma-\frac{1}{2}}\mbox{Im}[E^{0\ast}_{+}
(t)P_{k\sigma-\frac{3}{2}}]\nonumber\\
&&\mbox{}+2\sum_q V_q\mbox{Im}
(\sum_{\sigma^\prime}P_{k+q \sigma\sigma^\prime}^\ast
P_{k\sigma\sigma^\prime}
+\rho_{cc,k+q,-\sigma\sigma}\rho_{cc,k,\sigma -\sigma})\nonumber\\
\label{fecoh}
&&\mbox{}-g\mu_BB\mbox{Im}\rho_{cc,k,\sigma-\sigma}\ .
\end{eqnarray}
The first two terms describe the generation rates by the
polarized laser pulses. As mentioned before, the selection rule requires that
the optical transition can only happen between the conduction and the valence
bands with the same spin direction. This selection rule is enforced
by the Kronecker $\delta$-function $\delta_{\sigma\pm\frac{1}{2}}$.
The third term describes the exchange
interaction correction of the exciting laser by the electron-hole
attraction, thus it can be seen as a local field correction of the time
dependent bare Rabi frequency $dE^0_{\pm}(t)$. The last term describes the
spin flip and flop of electrons. It is noticed that if $\mbox{Im}
\rho_{cc,k,\sigma-\sigma}=0$, there is {\em no} spin flip and flop.
Therefore we call $\rho_{cc,k,\sigma-\sigma}$ spin coherence. It is
further noticed that $\rho_{cc,k,-\frac{1}{2}\frac{1}{2}}=
\rho_{cc,k,\frac{1}{2}-\frac{1}{2}}^\ast$. Similarly the coherent part
of the equation of motion for the hole distribution function is
written as
\begin{eqnarray}
&&\left.\frac{\partial f_{hk\sigma}}{\partial t}\right|_{\mbox{coh}}
=-2\sum_{q\sigma^\prime} V_q\mbox{Im}
(P_{k+q \sigma^\prime\sigma}P_{k\sigma^\prime\sigma}^\ast)\nonumber\\
\label{fhcoh}
&&\mbox{}\hspace{0.2cm}+d
\delta_{\sigma\frac{3}{2}}\mbox{Im}[E_{-}(t)^\ast P_{k\frac{1}{2}\sigma}]
+d\delta_{\sigma-\frac{3}{2}}\mbox{Im}[E_{+}(t)^\ast
P_{k-\frac{1}{2}\sigma}]\ .
\end{eqnarray}
One notices here that differing from the electron distribution function,
there is {\em no} terms like $g\mu_B B\mbox{Im}\rho_{vv,k,\sigma-\sigma}$
in the coherent part of equation of motion for the hole distribution
even if we do not neglect the contribution from $\rho_{vv,k,\sigma-\sigma}$.
Again this is due to the fact that $(S_v^x)_{\pm \frac{3}{2},
\pm\frac{3}{2}}=0$ and therefore the spin of hole cannot be flipped.
The scattering rates of $f_{ek\sigma}$ and $f_{hk\sigma}$ for
the Coulomb scattering in the Markovian limit are given by
\end{multicols}
\widetext
\begin{eqnarray}
&&\left.\frac{\partial f_{ek\sigma}}{\partial t}\right|_{\mbox{scat}}
=-2\sum_{j=e,h,k^\prime q\sigma^\prime}
V_q^22\pi\delta(\varepsilon_{ek-q}-
\varepsilon_{ek}+\varepsilon_{jk^\prime}-\varepsilon_{jk^\prime-q})
\Bigg\{(1-f_{ek-q\sigma})f_{ek\sigma}(1-f_{jk^\prime\sigma^\prime})
f_{jk^\prime-q\sigma^\prime}\nonumber\\
&&\mbox{}-f_{ek-q\sigma}(1-f_{ek\sigma})f_{jk^\prime\sigma^\prime}
(1-f_{jk^\prime-q\sigma^\prime})+\Big[\sum_{\sigma^{\prime\prime}}
\mbox{Re}(P_{k-q\sigma\sigma^{\prime
\prime}}P^\ast_{k\sigma\sigma^{\prime\prime}})\nonumber\\
&&\mbox{}+\mbox{Re}(\rho_{cc,k-q,\sigma-\sigma}\rho_{cc,k,-\sigma\sigma})\Big]
(f_{jk^\prime\sigma^\prime}-f_{jk^\prime-q\sigma^\prime})+
\sum_{\sigma^{\prime\prime}}
\mbox{Re}(P_{k^\prime\sigma^{\prime}\sigma^{\prime\prime}}P^\ast_{k^\prime
-q\sigma^{\prime}\sigma^{\prime\prime}})(f_{ek-q\sigma}-f_{ek\sigma})\Bigg\}
\nonumber\\
\label{fescat}
&&\mbox{}-2\sum_{k^\prime q \sigma^\prime}V_q^22\pi\delta(\varepsilon_{ek-q}-
\varepsilon_{ek}+\varepsilon_{ek^\prime}-\varepsilon_{ek^\prime-q})
\rho_{cc,k^\prime,\sigma^\prime-\sigma^\prime}
\rho_{cc,k^\prime-q,-\sigma^\prime\sigma^\prime}(f_{ek-q\sigma}-f_{e
k\sigma})\ ,
\end{eqnarray}
and
\begin{eqnarray}
&&\left.\frac{\partial f_{hk\sigma}}{\partial t}\right|_{\mbox{scat}}
=2\sum_{j=e,h,k^\prime q\sigma^\prime}
V_q^22\pi\delta(\varepsilon_{hk}-
\varepsilon_{hk-q}+\varepsilon_{jk^\prime-q}-\varepsilon_{jk^\prime})
\Bigg\{f_{hk-q\sigma}(1-f_{hk\sigma})(1-f_{jk^\prime-q\sigma^\prime})
f_{jk^\prime\sigma^\prime}\nonumber
\\
&&\mbox{}-f_{hk\sigma}(1-f_{hk-q\sigma})f_{jk^\prime-q\sigma^\prime}
(1-f_{jk^\prime\sigma^\prime})-\sum_{\sigma^{\prime\prime}}[
\mbox{Re}(P^\ast_{k-q\sigma^{\prime\prime}\sigma}
P_{k\sigma^{\prime\prime}\sigma})
(f_{jk^\prime\sigma^\prime}-f_{jk^\prime-q\sigma^\prime})
\nonumber
\\
&&\mbox{}+\mbox{Re}(P^\ast_{k^\prime-q\sigma^\prime\sigma^{\prime\prime}}
P_{k^\prime\sigma^\prime\sigma^{\prime\prime}})(f_{hk-q\sigma}-
f_{hk\sigma})]\Bigg\}+2\sum_{k^\prime q\sigma^\prime}
V_q^22\pi\delta(\varepsilon_{hk}-\varepsilon_{hk-q}
+\varepsilon_{ek^\prime-q}-\varepsilon_{ek^\prime})\nonumber
\\
\label{fhscat}
&&\mbox{}\hspace{1cm}\times\rho_{cc,k^\prime-q,\sigma^\prime-\sigma^\prime}
\rho_{cc,k^\prime,-\sigma^\prime\sigma^\prime}(f_{hk\sigma}
-f_{hk-q\sigma})\ .
\end{eqnarray}
One can easily prove from Eqs.\ (\ref{fescat}) and (\ref{fhscat}) that
$\sum_k\frac{\partial f_{ek\sigma}}{\partial t}|_{\mbox{scat}}=
\sum_k\frac{\partial f_{hk\sigma}}{\partial t}|_{\mbox{scat}}=0$.
This is because the Coulomb scattering does not change the total population
of each band.

The coherent time evolution of the interband polarization component is
given by
\begin{eqnarray}
\left.\frac{\partial}{\partial t}P_{k\sigma\sigma^\prime}\right|_
{\mbox{coh}}&=&-i\delta_{\sigma\sigma^\prime}(k)P_{k
\sigma\sigma^\prime}-\frac{i}{2}g\mu_BBP_{k-\sigma\sigma^\prime}
+\frac{i}{2}dE_{-}(t)[\delta_{\sigma\frac{1}{2}}
(1-f_{hk\frac{3}{2}})
-\rho_{cc,k,\sigma\frac{1}{2}}]\delta_{\sigma^\prime\frac{3}{2}}
\nonumber\\
&&\mbox{}+\frac{i}{2}dE_{+}(t)[\delta_{\sigma-\frac{1}{2}}
(1-f_{hk-\frac{3}{2}})
-\rho_{cc,k,\sigma-\frac{1}{2}}]\delta_{\sigma^\prime-\frac{3}{2}}
\nonumber\\
\label{pcoh}
&&\mbox{}-i\sum_qV_q\left[P_{k+q,\sigma\sigma^\prime}(1-f_{hk\sigma^\prime}
-f_{ek\sigma})
-P_{k+q,-\sigma\sigma^\prime}\rho_{cc,k,\sigma-\sigma}+
\rho_{cc,k+q,\sigma-\sigma}P_{k,-\sigma\sigma^\prime}\right]\ .
\end{eqnarray}
The first term gives the free evolution of the polarization components with
the detuning
\begin{equation}
\delta_{\sigma\sigma^\prime}(k)=\varepsilon_{ek}
+\varepsilon_{hk}-\Delta_0-\sum_qV_q(f_{ek+q\sigma}+f_{hk+q\sigma^\prime})
\end{equation}
and $\Delta_0=\omega-E_g$. $\Delta_0$ is the detuning of the center
frequency of the light pulses with respect to the unrenormalized
band gap. The second term in Eq.\ (\ref{pcoh})
describes the coupling of the optical coherence
$P_{k\sigma\sigma}$ with the forbidden optical coherences
$P_{k\sigma-\sigma}$ and $P_{k-\sigma\sigma}$ due to the
presence of the magnetic field $B$. This coupling makes
$P_{k\sigma-\sigma}$ and $P_{k-\sigma\sigma}$  not small enough
to be neglected in our calculation. The last term in Eq.\ (\ref{pcoh})
describes again the excitonic correlations. The coherent time evolution of
the spin coherence is given by
\begin{eqnarray}
\left.\frac{\partial}{\partial t}\rho_{cc,k,\sigma-\sigma}\right|_
{\mbox{coh}}&=&\frac{i}{2}d(\delta_{\sigma\frac{1}{2}}E_{-}(t)
P^\ast_{k-\sigma\frac{3}{2}}-\delta_{-\sigma\frac{1}{2}}E_{-}(t)^\ast
P_{k\sigma\frac{3}{2}})+\frac{i}{2}d(\delta_{\sigma-\frac{1}{2}}
E_{+}(t)P^\ast_{k-\sigma-\frac{3}{2}}-\delta_{-\sigma-\frac{1}{2}}
E_{+}(t)^\ast P_{k\sigma-\frac{3}{2}})\nonumber\\
&&\mbox{}+i\sum_q
V_q[(f_{ek+q\sigma}-f_{ek+q-\sigma})\rho_{cc,k,\sigma-\sigma}-
\rho_{cc,k+q,\sigma-\sigma}(f_{ek\sigma}-f_{ek-\sigma})\nonumber\\
\label{rhocoh}
&&\mbox{}+P_{k+q\sigma\sigma_1}P^\ast_{k-\sigma\sigma_1}-
P^\ast_{k+q-\sigma\sigma_1}P_{k\sigma\sigma_1}]+
\frac{i}{2}g\mu_BB(f_{ek\sigma}-f_{ek-\sigma})\ .
\end{eqnarray}
One notices from both the last terms of Eqs.\ (\ref{fecoh}) and (\ref{rhocoh})
that the spin coherence causes the electrons oscillating between the spin-up
and spin-down bands and in the mean time the unbalance between these two bands
feeds back to the spin coherence. It is further noted from the first
two terms of Eq.\ (\ref{rhocoh}) that pump pulse also generates the
spin coherence. However, as this process
is through the coupling to the forbidden transitions, its contribution
is negligible compared to the effect due to magnetic field [last term of
Eq. (\ref{rhocoh})] as will be shown in the later sections.

The dephasing of the interband polarization components is determined by
the following scattering:
\begin{eqnarray}
&&\left.\frac{\partial P_{k\sigma\sigma_0}}
{\partial t}\right|_{\mbox{scat}}=
\Bigg\{-\sum_{\stackrel{j=e,h}{k^\prime q\sigma^\prime}}
V_q^22\pi\delta(\varepsilon_{ek-q}
+\varepsilon_{hk}+\varepsilon_{jk^\prime}-\varepsilon_{jk^\prime-q}-\Delta_0)
\Big\{(P_{k,\sigma\sigma_0}-P_{k-q,\sigma\sigma_0})\Big[(1-
f_{jk^\prime\sigma^\prime})f_{jk^\prime-q\sigma^\prime}\nonumber
\\
&&\mbox{}-\sum_{\sigma^{\prime\prime}}
P_{k^\prime,\sigma^\prime\sigma^{\prime\prime}}
P^\ast_{k^\prime-q,\sigma^\prime\sigma^{\prime\prime}}\Big]+
(\rho_{cc,k-q,\sigma-\sigma}P_{k,-\sigma\sigma_0}+
f_{ek-q\sigma}P_{k\sigma\sigma_0}-f_{hk\sigma_0}P_{k-q\sigma\sigma_0})
(f_{jk^\prime\sigma^\prime}-f_{jk^\prime-q
\sigma^\prime})\Big\}\nonumber
\\
&&\mbox{}+\sum_{k^\prime q\sigma^\prime}V_q^22\pi\delta(\varepsilon_{ek-q}
+\varepsilon_{hk}+\varepsilon_{ek^\prime}-\varepsilon_{ek^\prime-q}
-\Delta_0)
(P_{k\sigma\sigma_0}-P_{k-q\sigma\sigma_0})\rho_{cc,k^\prime,\sigma^\prime
-\sigma^\prime}\rho_{cc,k^\prime-q,-\sigma^\prime\sigma^\prime}\Bigg\}
\nonumber\\
\label{pscat}
&&\mbox{}-\{k\leftrightarrow k-q,k^\prime\leftrightarrow k^\prime-q\}
-\frac{P_{k\sigma\sigma_0}}{T_2}\ .
\end{eqnarray}
Here $T_2$ is introduced phenomenologically to describe additional slower
scattering process such as carrier-phonon scattering. $\{k\leftrightarrow
k-q,k^\prime\leftrightarrow k^\prime-q\}$ in Eq.\ (\ref{pscat}) stands for
the same terms as in the previous $\{\}$ but with the interchanges
$k\leftrightarrow k-q$ and $k^\prime\leftrightarrow k^\prime-q$.
We point out here that all the scattering terms in Eqs.\ (\ref{fescat}),
(\ref{fhscat}) and (\ref{pscat}) are the contributions from the Coulomb
scattering. We did not include the EHSE scatterings as they are much weaker
in comparison with the Coulomb scatterings.

The Coulomb scattering contribution to the scattering term of the
spin coherence is:
\begin{eqnarray}
&&\left.\frac{\partial \rho_{cc,k\sigma-\sigma}}
{\partial t}\right|_{\mbox{scat}}^{\mbox{Coul}}
=\Bigg\{-\sum_{\stackrel{j=e,h}{qk^\prime\sigma^\prime}}
V_q^22\pi\delta(\varepsilon_{ek-q}
-\varepsilon_{ek}+\varepsilon_{jk^\prime}-\varepsilon_{jk^\prime-q})
\Big\{(f_{ek-q\sigma}\rho_{cc,k,\sigma-\sigma}+\rho_{cc,k-q,\sigma-\sigma}
f_{ek-\sigma}\nonumber
\\
&&\mbox{}+\sum_{\sigma^{\prime\prime}}P_{k-q\sigma\sigma^{\prime\prime}}
P^\ast_{k-\sigma-\sigma^{\prime\prime}})(f_{jk^\prime
\sigma^\prime}-f_{jk^\prime-q\sigma^\prime})+\rho_{cc,k,\sigma-\sigma}
\Big[(1-f_{jk^\prime\sigma^\prime})f_{jk^\prime-q\sigma^\prime}
-\sum_{\sigma^{\prime\prime}}
P_{k^\prime\sigma^\prime\sigma^{\prime\prime}}P^\ast_{k^\prime-q
\sigma^\prime\sigma^{\prime\prime}}\nonumber
\\
&&\mbox{}-\delta_{j=e}\rho_{cc,k^\prime,\sigma^\prime-\sigma^\prime}
\rho_{cc,k^\prime-q,-\sigma^\prime\sigma^\prime}\Big]-\rho_{cc,k-q,
\sigma-\sigma}\Big[f_{jk^\prime\sigma^\prime}(1-f_{jk^\prime-q\sigma^\prime})
-\sum_{\sigma^{\prime\prime}}P_{k^\prime\sigma^\prime
\sigma^{\prime\prime}}P^\ast_{k^\prime-q
\sigma^\prime\sigma^{\prime\prime}}\nonumber
\\
\label{rhocou}
&&\mbox{}-\delta_{j=e}\rho_{cc,k^\prime,\sigma^\prime-\sigma^\prime}
\rho_{cc,k^\prime-q,-\sigma^\prime\sigma^\prime}\Big]\Big\}\Bigg\}
-\{k\leftrightarrow k-q,k^\prime\leftrightarrow k^\prime-q\}\ .
\end{eqnarray}
\begin{multicols}{2}
\narrowtext
\noindent One can prove from Eq.\ (\ref{rhocou}) analytically that
\begin{equation}
\label{rea1}
\sum_k \left.\frac{\partial \rho_{cc,k\sigma-\sigma}}{\partial t}
\right|_{\mbox{scat}}^{\mbox{Coul}}=0\ .
\end{equation}
This can be easily seen as the second
half of Eq.\ (\ref{rhocou}) comes from the first half by  interchanging
$k\leftrightarrow k-q$ and $k^\prime\leftrightarrow k^\prime-q$. In the
sum over $k$ of Eq.\ (\ref{rea1}), one may perform the following
variable transformations: $k\rightarrow -k+q$ and $k^\prime\rightarrow
-k^\prime+q$ to the second half of Eq.\ (\ref{rhocou})
which make the second half just the same as the first one but with opposite
sign. Eq.\ (\ref{rea1}) indicates that the Coulomb scattering does not
contribute to the spin dephasing.

In order to study the dephasing of the spin coherence, we pick up the
contributions from EHSE scattering.  The ``direct'' EHSE contribution
is given by
\begin{eqnarray}
&&\left.\frac{\partial \rho_{cc,k\sigma-\sigma}}{\partial t}
\right|_{\mbox{scat}}^{\mbox{dEHSE}}\nonumber\\
&&\mbox{}=\Big\{-\frac{9}{16}\sum_{k^\prime q\sigma^\prime}
U_q^2 2\pi\delta(\varepsilon_{ek-q}-\varepsilon_{ek}
-\varepsilon_{hk^\prime}+\varepsilon_{hk^\prime-q})\nonumber\\
&&\mbox{}\times[(2-f_{ek\sigma}-f_{ek-\sigma})\rho_{cc,k-q,\sigma
-\sigma}f_{hk^\prime-q\sigma^\prime}(1-f_{hk^\prime\sigma^\prime})\nonumber
\\
&&\mbox{}+\rho_{cc,k,\sigma-\sigma}(f_{ek-q\sigma}+f_{ek-q-\sigma})
f_{hk^\prime-q\sigma^\prime}(1-f_{hk^\prime\sigma^\prime})\Big\}\nonumber
\\
\label{rhoehse}
&&\mbox{}+\{k\leftrightarrow k-q,k^\prime\leftrightarrow k^\prime-q\}\ .
\end{eqnarray}
The ``exchange'' EHSE contribution can be written as
\begin{eqnarray}
&&\left.\frac{\partial \rho_{cc,k\sigma-\sigma}}{\partial t}
\right|_{\mbox{scat}}^{\mbox{eEHSE}}\nonumber\\
&&\mbox{}=\Big\{-\frac{9}{16}\sum_{k^\prime q\sigma^\prime}
U_q^{\prime 2} 2\pi\delta(\varepsilon_{ek^\prime}+\varepsilon_{hk^\prime-q}
-\varepsilon_{ek}-\varepsilon_{hk-q})\nonumber\\
&&\mbox{}\times[(2-f_{ek^\prime\sigma}-f_{ek^\prime-\sigma})\rho_{cc,k,\sigma
-\sigma}f_{hk-q\sigma^\prime}(1-f_{hk^\prime-q\sigma^\prime})\nonumber
\\
&&\mbox{}+\rho_{cc,k^\prime,\sigma-\sigma}(f_{ek\sigma}+f_{ek-\sigma})
f_{hk-q\sigma^\prime}(1-f_{hk^\prime-q\sigma^\prime})\Big\}\nonumber
\\
\label{rhoehse1}
&&\mbox{}+\{k\leftrightarrow k^\prime\}\ .
\end{eqnarray}
It is noted that in  Eqs.\ (\ref{rhoehse}) and
(\ref{rhoehse1}), the second half of each equation shares the same sign
as the first half. Therefore $\sum_k \left.\frac{\partial
\rho_{cc,k\sigma-\sigma}}{\partial t}\right|_{\mbox{scat}}^{\mbox{dEHSE}}
\not=0$ and $\sum_k \left.\frac{\partial
\rho_{cc,k\sigma-\sigma}}{\partial t}\right|_{\mbox{scat}}^{\mbox{eEHSE}}
\not=0$. This tells us that EHSE contributes to the spin dephasing.
One further finds by comparing Eqs.\ (\ref{rhoehse}) and
(\ref{rhoehse1}) that besides the above mentioned $U_q^\prime\ll
U_q$, the energy phase space of the ``exchange'' EHSE, which is imposed by
the energy conservation, is quite limited compared to that of ``direct''
EHSE. All these indicate that the contribution of the ``exchange'' EHSE is
negligible in comparison with that of the ``direct'' EHSE.
We have further proved that there is {\em no} contribution to the scattering
term from the combination of Coulomb scattering and EHSE as $U_qV_q$ or
$U_q^\prime V_q$.
Therefore
\begin{eqnarray}
\left.\frac{\partial \rho_{cc,k\sigma-\sigma}}{\partial t}
\right|_{\mbox{scat}}&=&\left.\frac{\partial
\rho_{cc,k\sigma-\sigma}}{\partial t}\right|_{\mbox{scat}}^{\mbox{Cou}}
+\left.\frac{\partial \rho_{cc,k\sigma-\sigma}}{\partial t}
\right|_{\mbox{scat}}^{\mbox{dEHSE}}\nonumber\\
\label{rhoscat}
&&\mbox{}+\left.\frac{\partial \rho_{cc,k\sigma-\sigma}}{\partial t}
\right|_{\mbox{scat}}^{\mbox{eEHSE}}\ .
\end{eqnarray}

Equations\ (\ref{kin})-(\ref{rhoscat}) comprise
the complete set of kinetic
equations. It is noted that we only include EHSE in the scattering
term of the spin coherence. Its contribution to the optical coherences and
electron (hole) distributions are neglected as it's much smaller than the
contribution from the Coulomb scattering.\cite{comment} Moreover, the
electron-hole recombination is not included in our model as the time scale
for such effect is at least one order of magnitude longer than the time
scale of dephasing.

\subsection{Faraday rotation angle}

The FR angle can be calculated for two degenerate Gaussian
pulses with variable delay time $\tau$. The first pulse (pump) is circular
polarized, {\em eg.} $E^0_{\mbox{pump}}=E^0_-(t)$, and
travels in the direction ${\bf k}_1$. The
second pulse (probe) is linear polarized and is much weaker than the first
one, {\em eg.} $E^0_{\mbox{prob}}(t)=E^0_{\mbox{prob},-}(t-\tau)
+E^0_{\mbox{prob},+}(t-\tau)\equiv\chi[E^0_-(t-\tau)+E^0_+(t-\tau)]$ with
$\chi\ll 1$. The probe pulse travels in the ${\bf k}_2$ direction.

The FR angle is defined as\cite{linder,lu}
\begin{eqnarray}
\label{fr}
\Theta_F(\tau)&=&C\sum_{k}\int \mbox{Re}
\Big[\bar P_{k\frac{1}{2}\frac{3}{2}}(t)E^{0\ast}_
{\mbox{prob},-}(t-\tau)\nonumber\\
&&\mbox{}\hspace{1cm}-\bar P_{k-\frac{1}{2}-\frac{3}{2}}(t)E^{0\ast}_
{\mbox{prob},+}(t-\tau)\Big]dt\ ,
\end{eqnarray}
with $\bar P_{k\sigma\sigma}$ standing for the optical transition in
the prob direction, {\em i.e.} ${\bf k}_2$ direction. $C$ is a constant.

For the delay time $\tau$ is shorter than the optical dephasing time,
one has to  project the
optical transition $P_{k\sigma\sigma}$ to ${\bf k}_2$ direction.
One may use an adiabatic projection technique described in detail
in Ref.\ \onlinecite{bay}. This technique is suitable for optically
thin crystals, where the spatial dependence can be treated
adiabatically.\cite{koch} To do so, one replaces the single-pulse
envelope function in Eqs.\ (\ref{fecoh}) and (\ref{fhcoh})
by two delayed pulses $E^0_-(t)=E^0_{-}(t)e^{i\varphi}+
E^0_{\mbox{prob},-}(t-\tau)$ and $E^0_+(t)=E^0_{\mbox{prob},+}(t-\tau)$
with the relative phase $\varphi=({\bf k}_1-{\bf k}_2)
\cdot {\bf x}$ resulting from the different propagation directions.
The projection technique is used with respect to this phase.
However, when delay time $\tau$ is much longer than the optical dephasing,
the optical transition $P_{k\sigma\sigma}$ induced by the pump pulse has
already decayed to zero and therefore one may perform the calculation
with $\varphi\equiv 0$.

It is interesting to see from Eq.\ (\ref{fr}) that although the spin
coherence is determined by $\rho_{cc,k,\frac{1}{2}-\frac{1}{2}}$, it does
not appear directly in the final equation of the FR
angle. Instead, it affects the FR
angle through the optical transitions $P_{k\sigma\sigma}$. For
delay time $\tau$ much longer than the optical dephasing time, the optical
coherences $P_{k\sigma\sigma}$ induced directly by the pump pulse together
with the forbidden optical coherences $P_{k\sigma-\sigma}$  have already
been destroyed to zero. However, the spin coherence
induced by the same pump
pulse remains and makes the electrons oscillate between the spin-up
and down bands. This unbalance of population strongly affects the optical
transitions induced later by the probe pulse around $\tau$
and gives rise to the time evolution of Faraday rotation angle.

\section{Numerical results}

We perform a numerical study of the Bloch equations to study the spin
coherence of optically excited
electrons in an undoped insulating ZnSe/Zn$_{1-x}$Cd$_x$Se quantum well.
 As the main interest of the present paper is focused on the mechanism
of the spin dephasing, we will not perform a pump-probe computation to
calculate the FR angle Eq.\ (\ref{fr}) as it
requires extensive CPU
time and also as it has been calculated by Linder and Sham\cite{linder}
for the same set of Bloch equations but with relaxation time approximation
for all the dephasing. Instead, we will only apply a single pump
pulse and calculate the time evolutions of both the optical and
spin coherences together with the
electron and hh distributions after that pulse
under the carrier-carrier scattering. The dephasing
of the spin coherence is well defined by the incoherently-summed
spin coherence, $\rho(t)=\sum_k|\rho_{cc,k,\frac{1}{2}-\frac{1}{2}}(t)|$,
as well as the optical dephasing is described by the incoherently summed
polarization, $P_{\sigma\sigma}(t)=\sum_k|P_{k,\sigma\sigma}(t)|$.
The later was first introduced
by Kuhn and Rossi.\cite{kuhn} It is understood that both true dissipation
and the interference of many $k$ states may contribute to the decay.
The incoherent summation is therefore used to isolate the irreversible decay
from the decay caused by interference. From $\rho(t)$ and $P_{\sigma
\sigma}(t)$ one gets the true irreversible dephasing of spin and optical
coherences respectively.
\begin{figure}[htb]
\vskip -1.cm
\psfig{figure=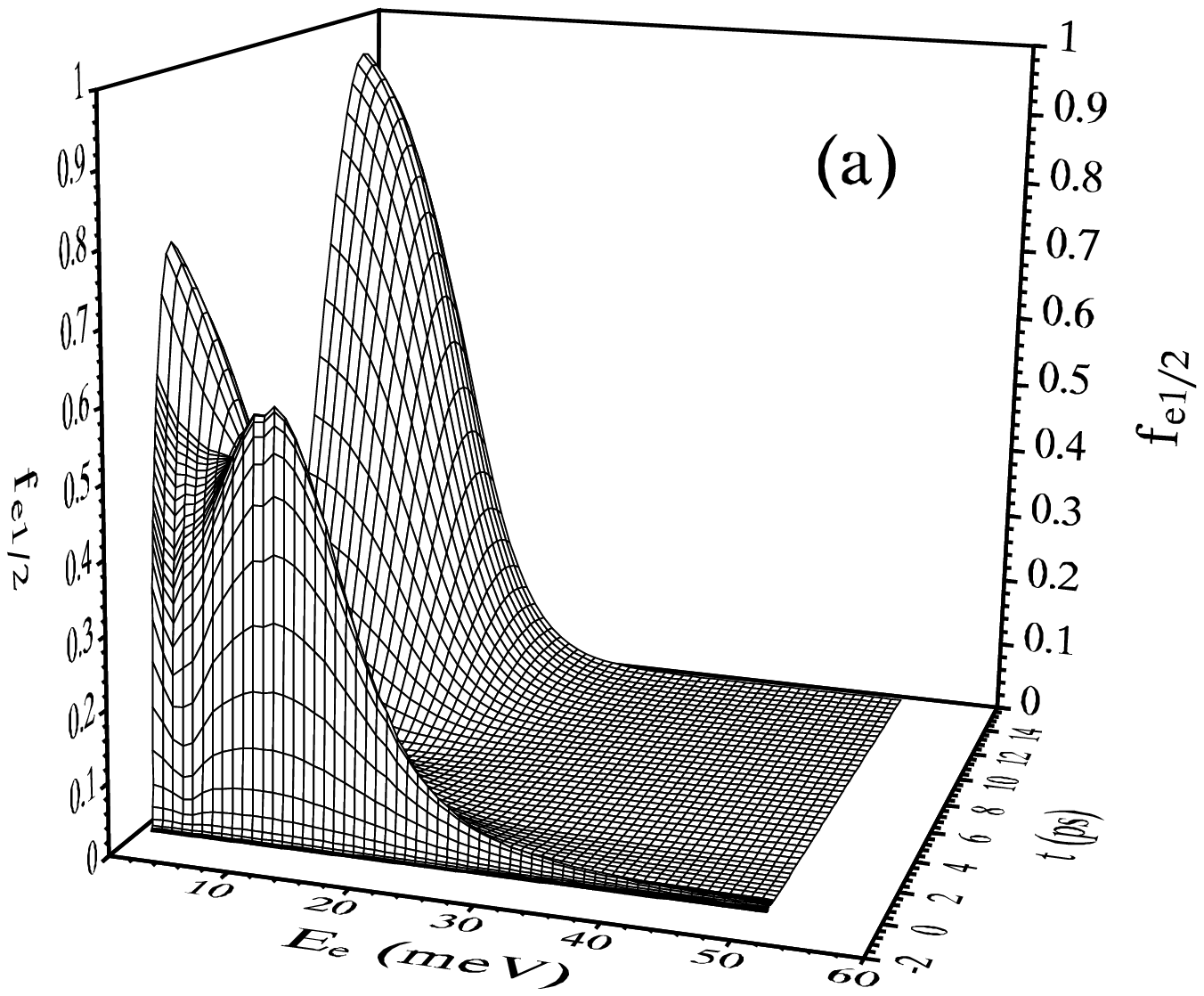,width=10.cm,height=8.5cm,angle=0}
\vskip -2.cm
\psfig{figure=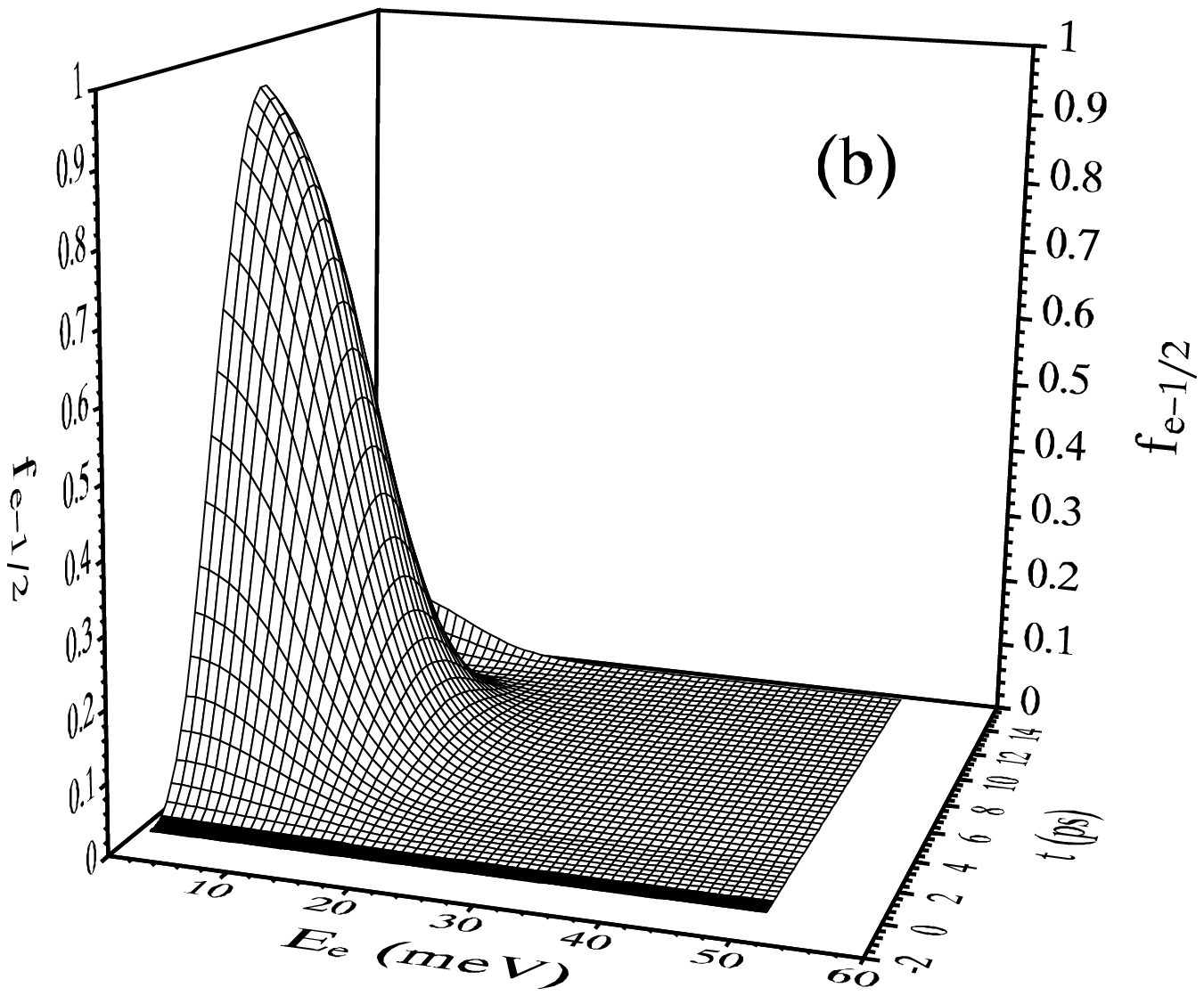,width=10.cm,height=8.5cm,angle=0}
\vskip -2.cm
\psfig{figure=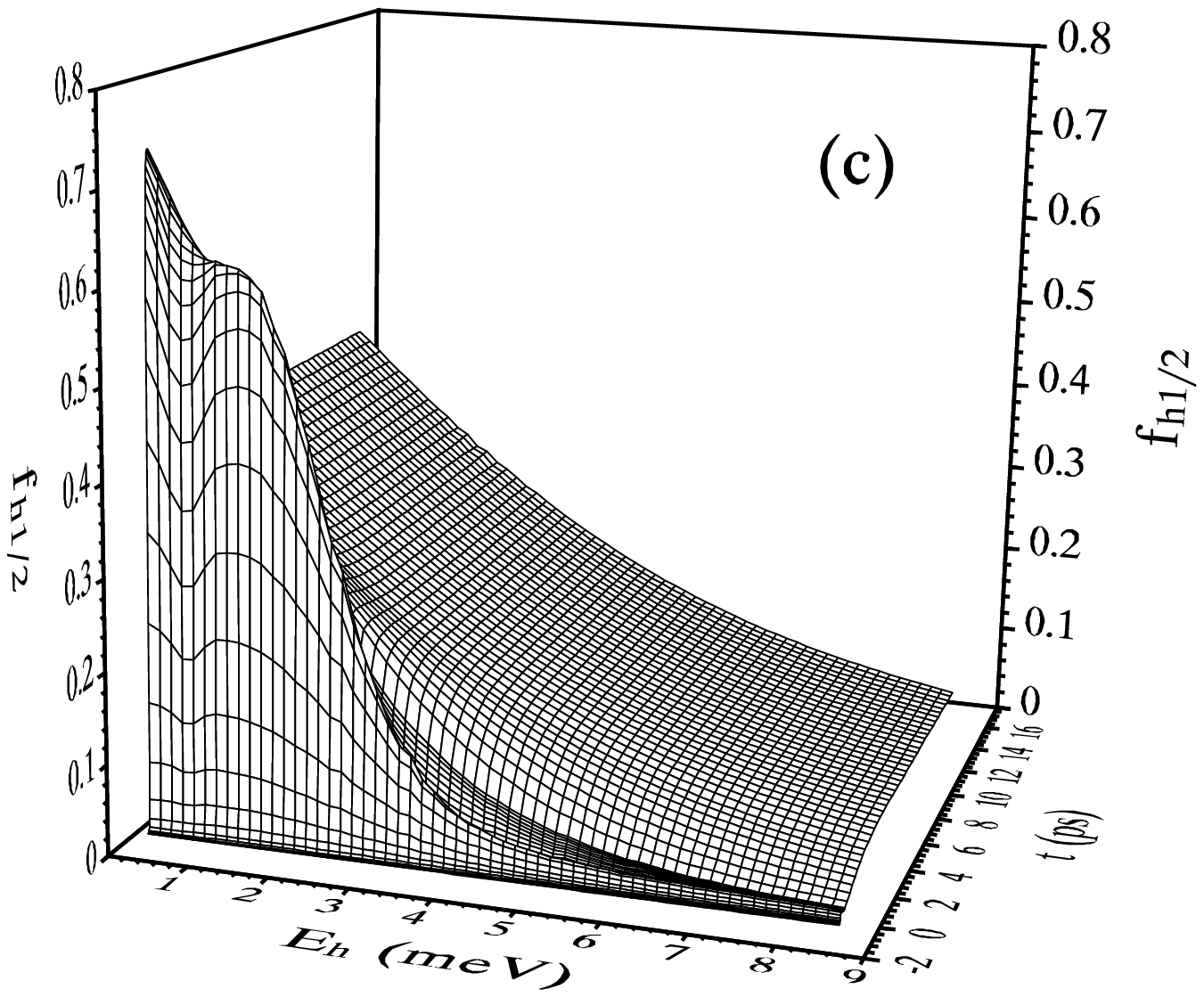,width=10.cm,height=8.5cm,angle=0}
\vskip -0.5cm
\caption{Electron distributions vs. time $t$ and electron energy $E_e$
for the spin 1/2 (Fig.\ 1(a)) and $-1/2$ (Fig.\ 1(b)) and hh distribution
vs. $t$ and hh energy $E_h$ (Fig.\ 1(c)).}
\end{figure}

The material parameters in our calculation are taken from the experimental
data with effective mass $m_e=0.152m_0$ for electron\cite{pri} and
$m_h=6m_e$ for hh.\cite{lan} Exciton Rydberg $E_R=19$\ meV and the $g$-factor is take
as 1.3. This $g$-factor is larger than what reported in the
experiment\cite{kikk1} (1.1) by measuring the beat frequency of
the FR angle. The reason will be shown clearly in
later subsections. $T_2$ is taken as 10\ ps for all the calculations.
We choose a left circular polarized Gaussian pulse
$E_-^0(t)$ with the width $\delta t=100$\ fs and detuning $\Delta_0=
4.275$\ meV. The total density excited by this pulse is $3\times
10^{11}$\ cm$^{-2}$, which is the same as what reported
in the experiment.\cite{kikk1}
$E_+^0(t)\equiv 0$. Under such a pulse, one can see immediately from the
kinetic equations that only hh with spin $\frac{3}{2}$ can be excited and
$f_{hk-\frac{3}{2}}\equiv 0$. Moreover, $P_{k-\frac{1}{2}-\frac{3}{2}}=
P_{k\frac{1}{2}-\frac{3}{2}}\equiv 0$. Therefore, we only need to determine
the electron distribution functions $f_{ek\sigma}$ ($\sigma=\pm 1/2$),
the hh distribution
functions $f_{hk\frac{3}{2}}$, interband polarizations (optical coherences)
$P_{k\frac{1}{2}\frac{3}{2}}$ and $P_{k-\frac{1}{2}\frac{3}{2}}$, and
spin coherences $\rho_{cc\frac{1}{2}-\frac{1}{2}}$. We will solve the
Bloch equations with only Coulomb scattering and with both Coulomb scattering
and EHSE scatterings respectively.
\begin{figure}[htb]
\vskip -1.cm
\psfig{figure=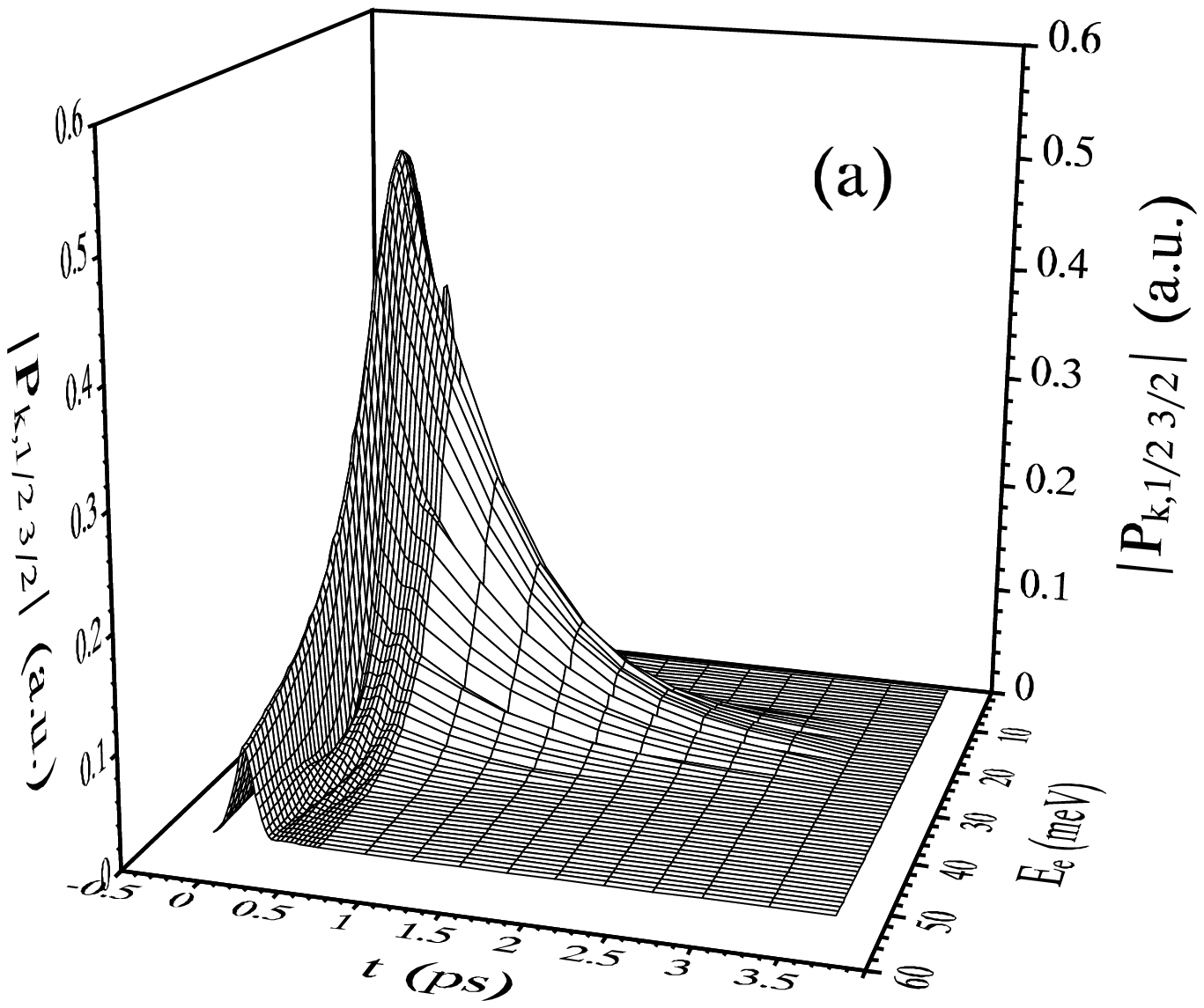,width=10.cm,height=8.5cm,angle=0}
\vskip -2.cm
\psfig{figure=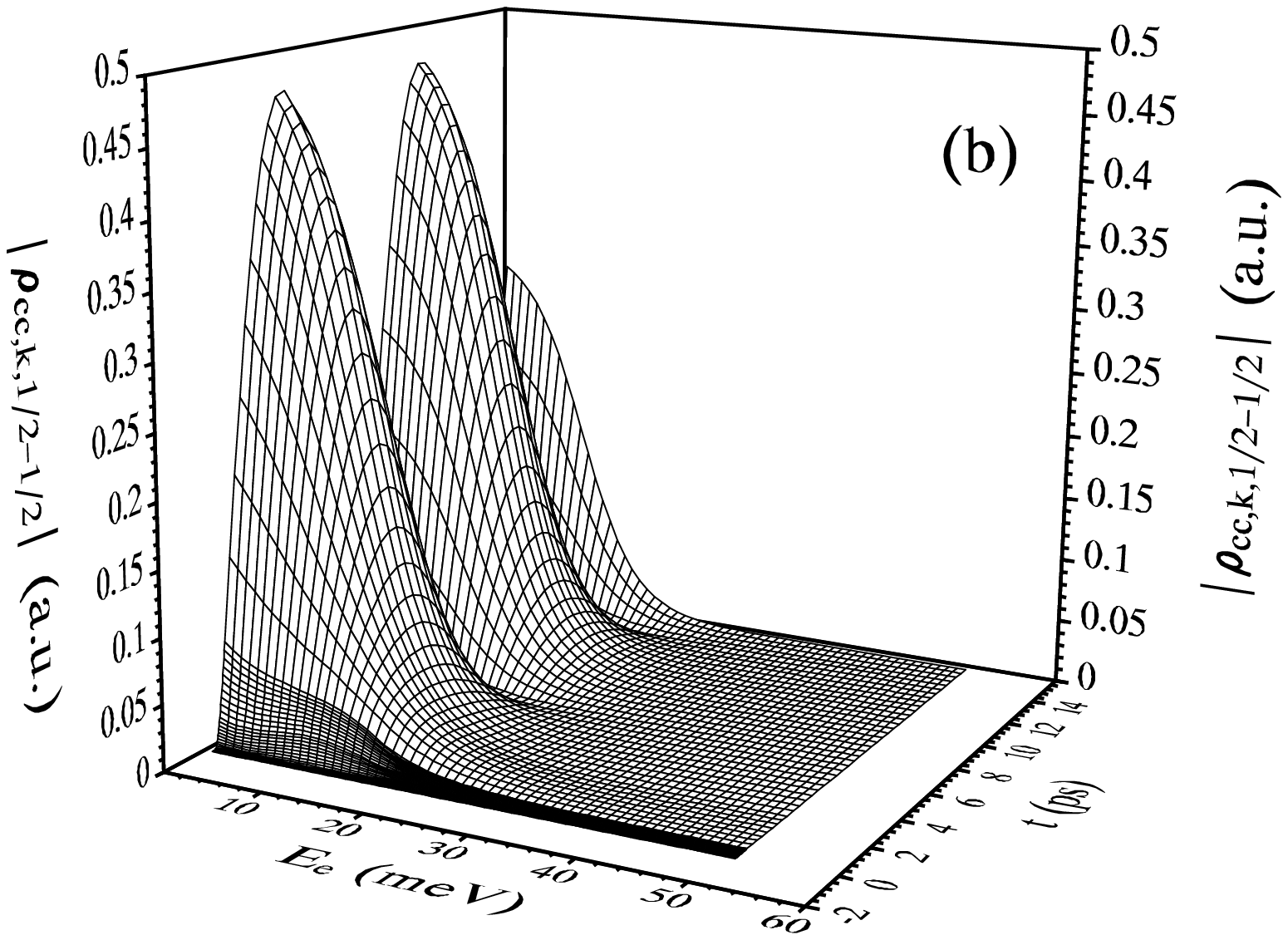,width=10.cm,height=8.5cm,angle=0}
\vskip -0.5cm
\caption{Absolute amounts of optical
coherence $|P_{k\frac{1}{2}\frac{3}{2}}(t)|$ [Fig.\ 2(a)]
and spin coherence $|\rho_{cc,k,\frac{1}{2}-\frac{1}{2}}(t)|$ [Fig.\ 2(b)]
as functions of $t$ and electron energy
$E_e$. Note the time scale of optical coherence is much shorter than the spin
coherence.}
\end{figure}

\subsection{Coulomb scattering}

We first discuss the Boltzmann kinetics under the Coulomb scattering
with the statically screened instantaneous potential approximation:
\begin{equation}
\label{vq}
V_q=\frac{2\pi e^2}{\epsilon_0(q+\kappa)}\ .
\end{equation}
The inverse screening length $\kappa$ is expressed as\cite{hk,haug}
\begin{equation}
\kappa(t)=\frac{m_e e^2}{\epsilon_0}\sum_\sigma[f_{ek=0\sigma}(t)+
\frac{m_h}{m_e}f_{hk=0\sigma}(t)]\ .
\end{equation}
It is noted that this static screening model, although simplifying numerical
calculation significantly, overlooks the plasmon-mode
contributions and overestimates the electron screening.
\begin{figure}[htb]
\psfig{figure=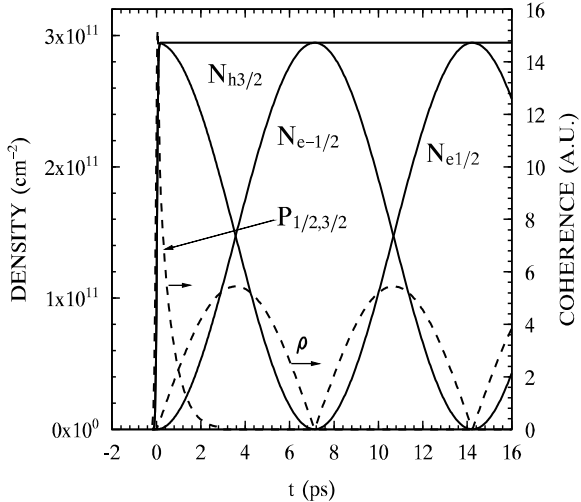,width=8.5cm,height=7.5cm,angle=0}
\caption{Total densities of each spin band $N_{ek\sigma} (t)$
for electron and $N_{hk\frac{3}{2}} (t)$ for hh (solid curves)
together with the
incoherently summed polarization $P_{\frac{1}{2}\frac{3}{2}}$ and
spin coherence $\rho(t)$ (dashed curves) are plotted against time $t$
for $B=4$\ T. Note the scale of the coherences is on the right side
of the figure.}
\end{figure}

We first show in Fig.\ 1 the electron and hole distribution functions
$f_{ek\sigma}(t)$ ($\sigma=\pm 1/2$) and $f_{hk\frac{3}{2}}(t)$
versus $t$ and electron energy $\varepsilon_{ek}$ (for electron
distributions) or hh energy $\varepsilon_{hk}$ (for
hole distribution) for $B=4$\ T. In the initial
time one observes a small peak around 10\ meV for $f_{ek\frac{1}{2}}$
in Fig.\ 1(a). This peak is
the effect of the pump pulse and the strong Hartree correction Eq.\ (10).
The similar peak can also be observed
for the hh distribution function in Fig.\ 1(c).
At later times the carriers relax to the low energy states and
$f_{ek\frac{1}{2}}$ reaches its first highest peak at 1.4\ ps.
Around 7\ ps, the distribution function of spin-up band reaches the
valley of zero values and the spin-down band,
in the mean time, arrives at its peak.
This indicates that electrons in the spin-up band evolve into the spin-down
band. After that electrons start to move back to the spin-up band and at about
14\ ps $f_{ek\frac{1}{2}}$ reaches its second highest peak and
$f_{ek-\frac{1}{2}}$ reaches its valley. The reason that the second
highest peak is a little higher than the first one is because there are more
electrons at higher energy states at initial times due to the
position of the pump pulse.  This oscillation keeps on without
any decay. The distribution for hh relaxes into the Fermi-like distribution
after a few picoseconds and remains unchanged. In Fig.\ 2
the absolute value of the optical coherence, $|P_{k,\frac{1}{2}
\frac{3}{2}}(t)|$, and the absolute value of the spin coherence,
$|\rho_{cc,k,\frac{1}{2}-\frac{1}{2}}(t)|$, are plotted as
functions of $t$ and electron energy.
It is seen from Fig.\ 2(a) that the optical coherence decays very
quickly and within first few picoseconds it
has already totally disappeared. Nevertheless, the spin coherence does not
decay at all. The second peak is of the same height as the first one.

The incoherently summed polarization, $P_{\frac{1}{2}\frac{3}{2}}(t)$, and
the incoherently summed spin coherence, $\rho(t)$,
are plotted as dashed curves in Fig.\ 3. The total densities of
each spin band $N_{e\sigma}(t)=\sum_kf_{ek\sigma}(t)$ for electron
and $N_{h\frac{3}{2}}(t)=\sum_kf_{hk
\frac{3}{2}} (t)$ for hh are also plotted as solid curves in the same figure.
It is seen from the figure that the optical coherence injected by the
pump pulse is about 3 times larger than the spin coherence. However, this
coherence is strongly dephased by the Coulomb scattering and is totally
gone within the first few picoseconds. It is further shown from the
figure that the spin coherence exhibits beating which
does not decay at all. The electron densities of spin-up and
down bands oscillate between zero and the total excitation. These
results confirm that pure Coulomb scattering does not contribute to
the spin dephasing. We further find that
the frequency of the oscillation is mainly determined by the Zeeman
split $g\mu_BB$, but red-shifted by the Hartree-Fock terms in
Eq.\ (\ref{rhocoh}). The reduced
$g$-factor resulting from the oscillation frequency is 1.25.
It is interesting to see from the
figure that the maximum value of $|N_{e\frac{1}{2}}-N_{e-\frac{1}{2}}|$
occurs when the (incoherently summed) spin coherence is zero. The forbidden
optical coherence is about 30 times smaller than the optical coherence and
decays similarly as the optical coherence. It plays an insignificant role
in this problem.

\subsection{EHSE}

Now we include the contribution of EHSE [Eqs.\ (\ref{rhoehse}) and
(\ref{rhoehse1})]. As said before that the contribution from ``exchange''
EHSE is negligible, we therefore only include the ``direct'' one
[Eq.\ (\ref{rhoehse})] here. We are lacking
information of the matrix elements $U_q$ which requires a detailed  band
structure calculation. For the sake of simplicity and also for
comparison with the effect of Coulomb scattering, in this study we assume
it is the same as $V_q$ but with a phenomenological
prefactor $4\sqrt{\cal F}/3$. We take
${\cal F}=0.015$ so that in the scattering term,
Eq.\ (\ref{rhoehse}), the matrix
element of EHSE is two orders of magnitude smaller than that of the
Coulomb scattering. This number is taken so that the spin dephasing time
is in agreement with the experiment.\cite{kikk1} We have also
performed numerical calculation by taking only the ``exchange''
EHSE and found in order to get the same spin dephasing, the prefactor
has to be 0.1 which is one order of magnitude larger than the present case.
This big prefactor is understood due to the effect of the limitation
of the energy phase space discussed before.
\begin{figure}[htb]
\vskip -1.cm
\psfig{figure=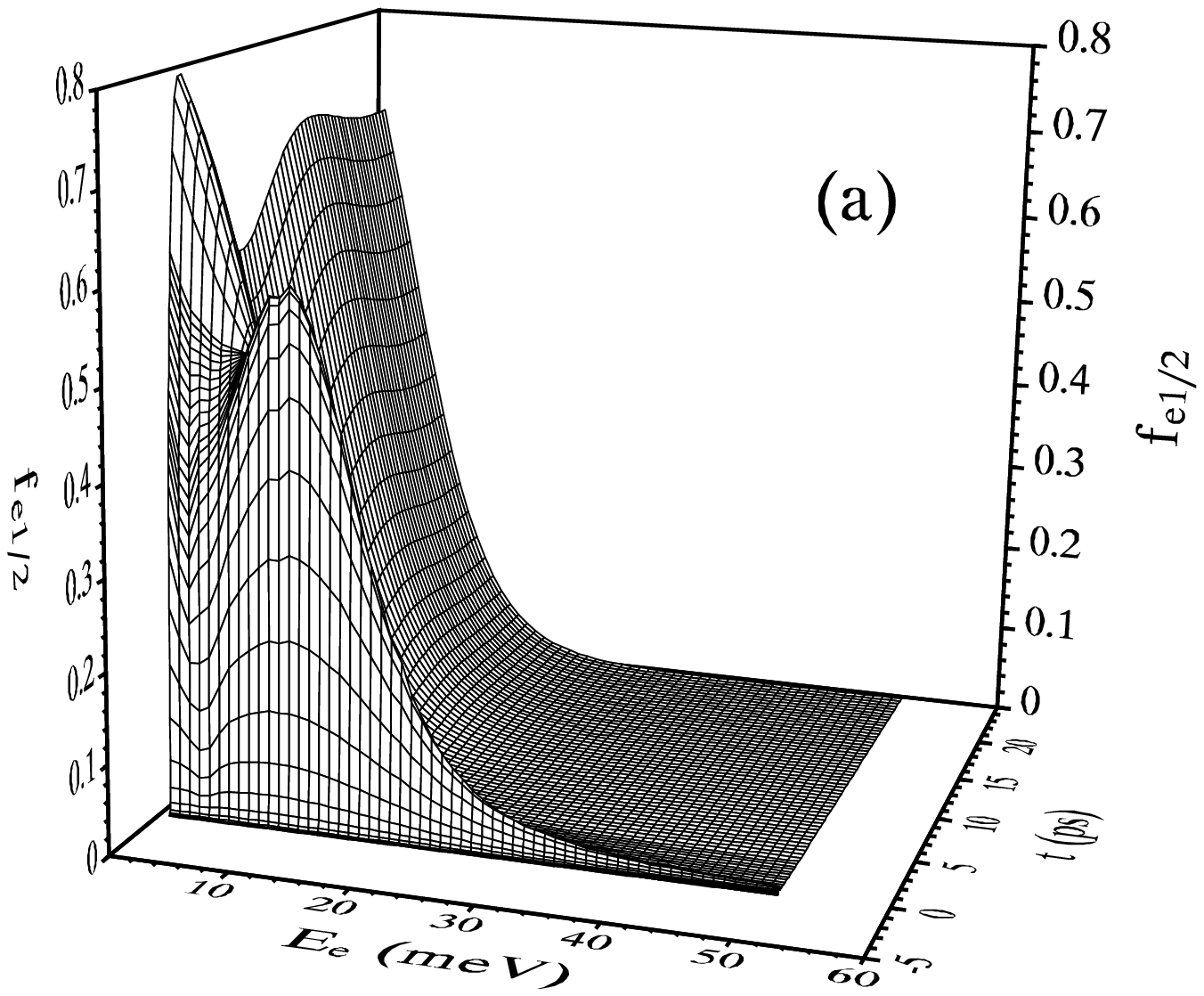,width=10.cm,height=8.5cm,angle=0}
\vskip -2.cm
\psfig{figure=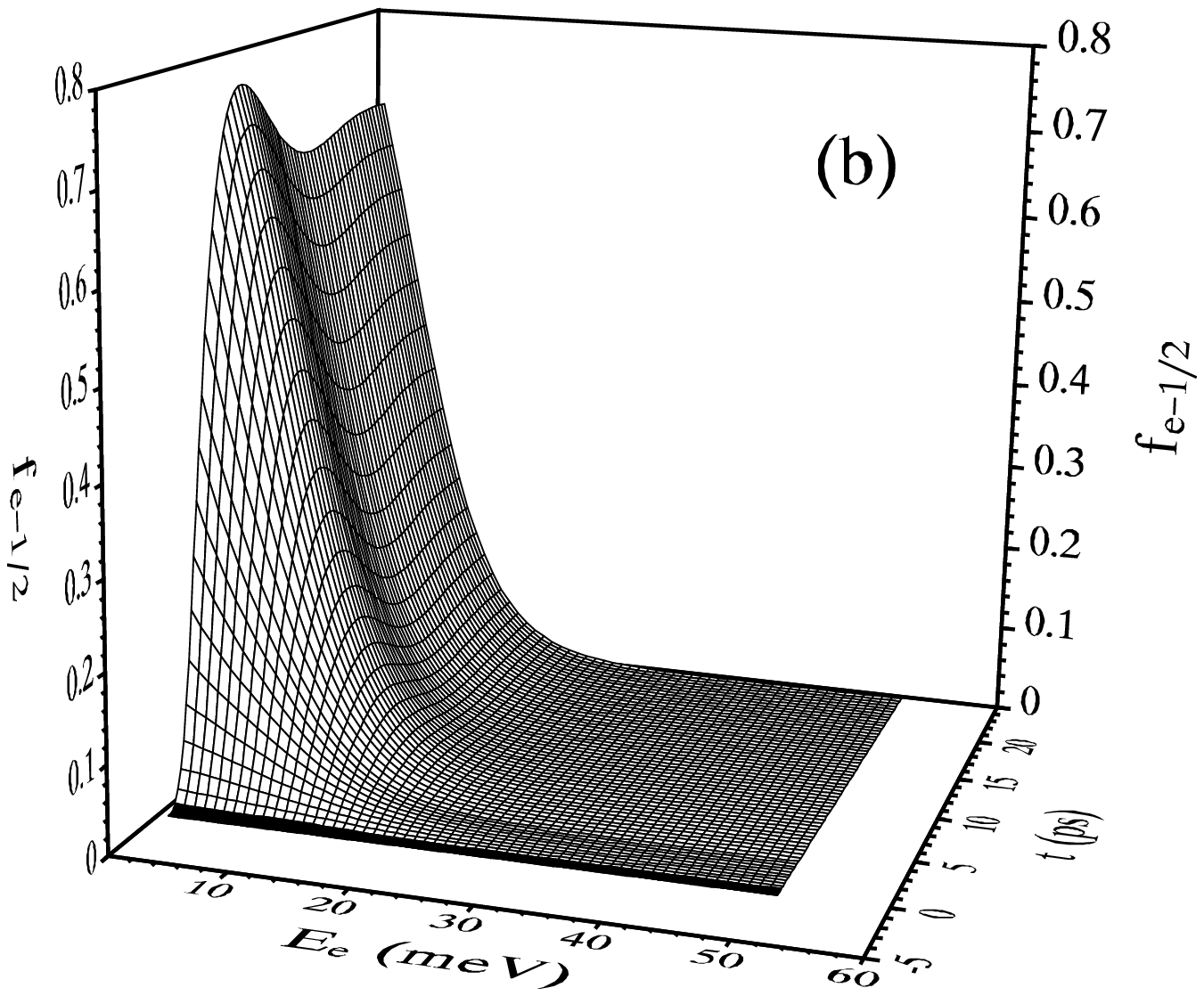,width=10.cm,height=8.5cm,angle=0}
\vskip -0.5cm
\caption{Electron distributions vs. time $t$ and electron energy $E_e$
for the spin 1/2 (Fig.\ 4(a)) and $-1/2$ (Fig.\ 4(b))
for $B=4$\ T. Effects of EHSE are included.}
\end{figure}

The electron distribution functions $f_{ek\sigma}(t)$ ($\sigma=\pm 1/2$)
versus $t$ and electron energy $\varepsilon_{ek}$
are plotted in Fig.\ 4 for $B=4$\ T. The hh distribution function
$f_{hk\frac{3}{2}}(t)$ versus $t$ and hh energy $\varepsilon_{hk}$
remains unchanged from Fig.\ 1(c) after the inclusion of EHSE.
In the initial time one observes again a small peak around 10\ meV for
$f_{ek\frac{1}{2}}$
in Fig.\ 4(a) due to the effect of pump detuning.
A similar peak can also be observed
for the hh distribution function as in Fig.\ 1(c). Again one observes that
the carriers relax at later times to the low energy states and all the
distributions show the Fermi-like distributions. Differing from the
case with only Coulomb scattering, there are only small
oscillations of electrons between spin 1/2 and $-1/2$ bands in the later
time. The hh distribution remains unchanged after the first few
picoseconds as before. The absolute value
of optical coherence, $|P_{k\frac{1}{2}\frac{3}{2}}(t)|$, versus
$t$ and electron energy is also unchanged from Fig.\ 2(a) after inclusion
of EHSE. Therefore, as before, the optical coherence
decays very quickly and within a few picoseconds it
has totally disappeared.
However, EHSE makes big change to the absolute value
of the spin coherence as can be seen in Fig.\ 5 where
$|\rho_{cc,k,\frac{1}{2}-\frac{1}{2}}(t)|$ is plotted
as a function of $t$ and electron energy.
It is seen that the spin coherence lasts much
longer and one can see a smaller second peak around 10.5\ ps and a
much smaller third peak around 18.4\ ps. Comparing with
Fig.\ 2(b), one can see the effect of strong spin-dephasing by EHSE
as the second peak is much lower than the first one. For $t>20$\ ps, the spin
coherence has almost gone. One more point needs to be addressed here is,
one can see from Fig.\ 5 that there is a very small bump
around 10\ meV at initial time. A similar bump can also be seen in
Fig.\ 2(b). These bumps are the effects of the pump pulse described before
after Eq.\ (\ref{rhocoh}). One can see that they are much smaller
compared to the effects caused by the magnetic field. This
also justifies the assumption we made before that hh-hh spin coherence
can be neglected in this investigation.
\begin{figure}[htb]
\vskip -1.cm
\psfig{figure=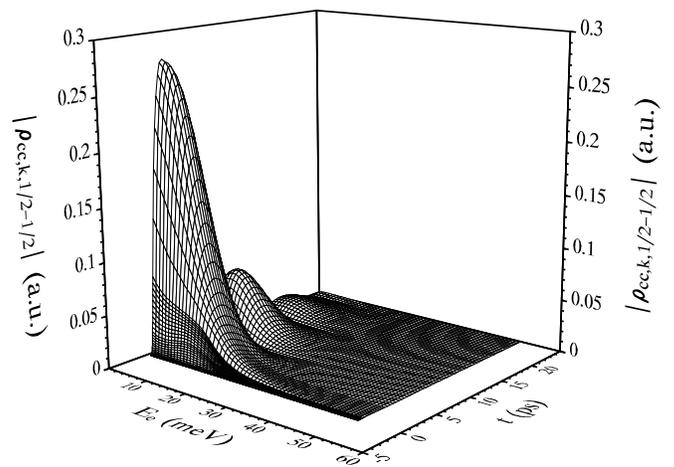,width=10.cm,height=8.5cm,angle=0}
\vskip -0.5cm
\caption{Absolute amounts of
spin coherence $|\rho_{cc,k,\frac{1}{2}-\frac{1}{2}}(t)|$
as functions of $t$ and electron energy $E_e$. Effects of
EHSE are included. $B=4$\ T. }
\end{figure}

The incoherently summed polarization, $P_{\frac{1}{2}\frac{3}{2}}(t)$, and
the incoherently summed spin coherence, $\rho(t)$,
are plotted as dashed curves in Fig.\ 6. In order
to have the effect of spin coherence more pronounced, we plot in the same
figure also $4.5\times\rho (t)$ as dash-dotted curve. The total densities of
each spin band $N_{e\sigma}(t)$ and $N_{h\frac{3}{2}}(t)$
are also plotted as solid curves in Fig.\ 6.
It is seen from the figure that the optical coherence injected by the
pump pulse is about 4.5 times larger than the spin coherence. However, this
coherence is strongly destroyed by the Coulomb scattering within the first
few picoseconds. This is part of the reason of the fast decay in
the FR angle for the first few picoseconds in
the experiment.\cite{kikk1} As seen
in Eq.\ (\ref{fr}), the FR angle is proportional to the
``total'' optical transitions. For the first few picoseconds, the optical
transitions are induced by both the pump pulse and the much weaker probe
pulse, and the fast decay of the initial pump pulse-induced optical
transition gives rise to the strong decay of the FR angle.
\begin{figure}[htb]
\psfig{figure=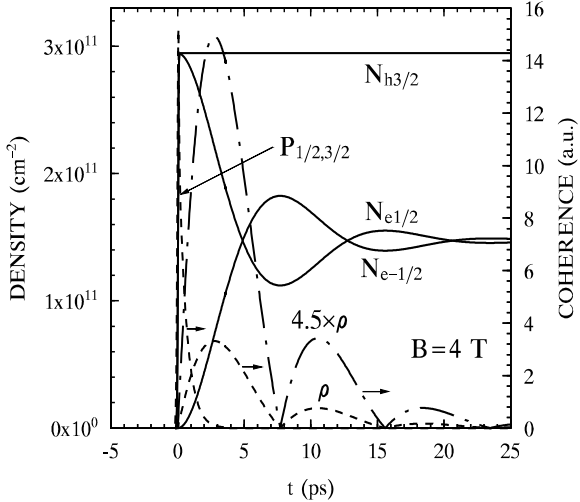,width=8.5cm,height=7.5cm,angle=0}
\caption{Total densities of each spin band $N_{ek\sigma} (t)$
for electron and $N_{hk\frac{3}{2}} (t)$ for hh (solid curves)
together with the
incoherently summed polarization $P_{\frac{1}{2}\frac{3}{2}}$ and
spin coherence $\rho(t)$ (dashed curves) are plotted against time $t$ for
$B=4$\ T. The dash-dotted curve is $4.5\rho(t)$. Effects of EHSE are
included. Note the scale of
coherences is on the right side of the figure.}
\end{figure}

It is further shown from the figure that the electron densities of spin
up and down bands oscillate with the amplitude of the oscillation
decaying. The circularly polarized pump pulse first
pumps electrons into the spin-up CB from the spin-up hh band. Therefore
electrons first occupy the spin-up CB band and leave behind hhs in
the spin-up VB. Therefore
$N_{e\frac{1}{2}}$ and $N_{h\frac{3}{2}}$ fast rise to
$3\times 10^{11}$\ cm$^{-2}$
within the time scale of the pump pulse.
Then due to the magnetic field, electrons
in the spin-up band start to go to the spin-down band. This makes $N_{e
\frac{1}{2}}$ decreases and $N_{e-\frac{1}{2}}$ rises. In the mean time,
the unbalance in populations also serves as pump field to the spin coherence.
After $t$ is around
5\ ps, the electron population in the spin-down band surpasses that in the
spin-up band. Without dephasing, this oscillation may keep going on as
shown in the previous section. However, the spin dephasing makes these two
populations finally merge. From the figure, one can see that for $t>20$\ ps,
the difference is already negligible compared to the oscillations
before and $\rho$ also decays to zero. This oscillation is also
shown in the experiment through the FR
angle as beatings for the same magnetic field.\cite{kikk1}
All these results confirm what we proposed in Sec.\ II, that the
spin dephasing is caused by EHSE and the electron spin coherence is
represented by $\rho$. Moreover, one also finds that
besides the effect of spin dephasing, EHSE
also changes the oscillation frequency and hence the
beating frequency of the FR angle. By comparing the period of
the oscillations in Figs.\ 3 and 6, one finds the period increases by about
1.34\ ps in the later case. This means that the EHSE further red-shifts the
spin splitting. Therefore, the $g$ factor reported in the
experiment\cite{kikk1} by measuring the frequency of the beating
of the FR angle is the {\em effective} $g$ factor.
It is seen from the figure that the period of the oscillation is
$T=15.58$\ ps. As $2\pi/T=g_{\mbox{eff}}\mu_BB$, the effective $g$-factor
$g_{\mbox{eff}}$ is therefore 1.14, in agreement with the what reported
in the experiment as 1.1.\cite{kikk1}
Again from the Figure one observes that the maximum value of
$|N_{e\frac{1}{2}}-N_{e-\frac{1}{2}}|$
occurs when the (incoherently summed) spin coherence is zero.
Our calculation shows again that the forbidden optical coherence is
insignificant in this problem.
\begin{figure}[htb]
\psfig{figure=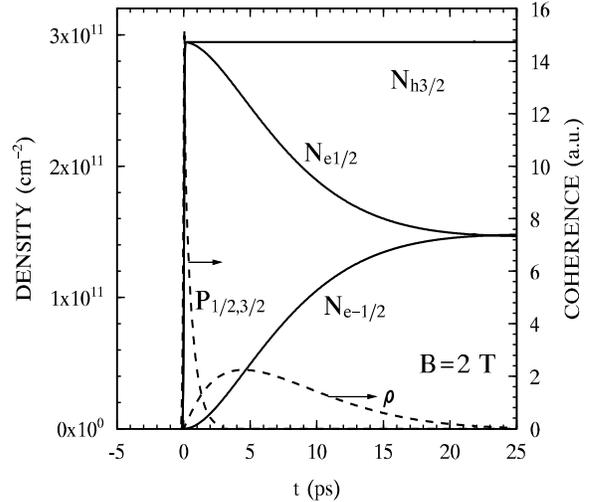,width=8.5cm,height=7.5cm,angle=0}
\caption{Total densities of each spin band $N_{ek\sigma} (t)$
for electron and $N_{hk\frac{3}{2}} (t)$ for hh (solid curves)
together with the
incoherently summed polarization $P_{\frac{1}{2}\frac{3}{2}}$ and
spin coherence $\rho(t)$ (dashed curves) are plotted against time $t$ for
$B=2$\ T. Effects of EHSE are included.
Note the scale of coherences is on the right side of the figure.}
\end{figure}

In order to compare with $B=2$\ T case, we plot in Fig.\ 7 the
incoherently summed polarization $P_{\frac{1}{2}\frac{3}{2}}(t)$ and
spin coherence $\rho(t)$ as function of time $t$, together with
the total densities of each spin band $N_{e\sigma}(t)$ and
$N_{h\frac{3}{2}}(t)$. One finds the spin coherence decays at
the same rate as in the $B=4$\ T case but without any
beating. So does the electrons in the spin bands. This confirms the
finding in the experiment that there is no beating in the
Faraday rotation angle.\cite{kikk1}

\section{Conclusion and Discussion}

In conclusion, we have performed theoretical studies of the kinetics of
the spin coherence of optically excited electrons in an undoped
insulating ZnSe/Zn$_{1-x}$Cd$_x$Se quantum well under moderate magnetic
fields in the Voigt configuration. Based on a two spin-band model in both
the CB and the VB, we build the kinetic equations combined with intra-
and interband Coulomb scattering and interband EHSE in the Boltzmann limit.
We include all the coherences induced directly by the laser pulse--- optical
transitions--- and indirectly through the effect of the magnetic
field--- electron-electron spin coherence and forbidden
optical coherence--- for the electron
and hh in our model. The hh-hh spin coherence $\rho_{vv,k,\frac{3}{2}
-\frac{3}{2}}$ is neglected in our present investigation because
it can not be induced by the effect of the magnetic field,
but only through the pump pulse coupled to the forbidden transitions
and is therefore much smaller than the other coherences.
We separate the spin coherence from the well known optical coherences and
study the effects of Coulomb scattering and EHSE on all the coherences.
We find that the Coulomb scattering makes strong dephasing of the optical
coherence and forbidden optical coherence. However, it
does not contribute to the spin dephasing at all.
EHSE is the main mechanism leads to the spin decoherence.
We numerically solve the kinetic equations for two different magnetic fields.
We find that the beating in the Faraday rotation
angle is basically determined by the electron Zeeman splitting
$g\mu_BB$, however, with a red shift from the
Coulomb Hartree-Fock contribution and EHSE effect. The forbidden optical
coherence is found of marginal importance in this problem.
The matrix element of the Coulomb scattering is taken as statically
screened Coulomb potential and the matrix element of EHSE is assumed
the same as the Coulomb scattering with a phenomenological prefactor.
By fitting this prefactor with the dephasing time in the
experiment,\cite{kikk1} our theory can well explain the experiment data
for two different magnetic fields.
A first principal investigation of the EHSE
scattering matrix element is definitely important for a more
thorough understanding of the spin dephasing.\cite{ehse,single}

For the $n$-doped material, things are quite different from the
undoped case discussed in this paper. Due to the presence of large
numbers of doped electrons (about one order of magnitude larger
than the optically excited carriers) in the CB, the lifetime of holes is
therefore quite short and is measured smaller than 50\ ps compared to
the lower limit of 100\ ps for the insulating sample.\cite{kikk1} The
experiments found that the spin dephasing for the doped sample is three
orders of magnitude longer than the undoped sample, but with a fast
dephasing (losing about 50\% coherence) within the first 50\ ps.
The fast dephasing in the initial times can be well understood by the EHSE
discussed above, but modified with the fast decreasing of hh population.
Nevertheless, the mechanism of the
spin dephasing discussed above cannot be applied to the
doped case after 50\ ps as shown from Eq.\ (\ref{rhoehse}) that the
scattering of EHSE is proportional to the hole distribution and therefore its
contribution to the spin dephasing decreases with the recombination of
electron and hole.
However,  for the $n$-doped case,  it is also possible for EHSE to
contribute to the spin dephasing. As the timescale of
the spin dephasing for the $n$-doped sample is much longer than the
insulating sample, the coupling between the hh and the light hole
can not be neglected and the spin of
hh also experiences the precession.\cite{marie}
Therefore, the hh-hh spin coherence
$\rho_{vv,k,\frac{3}{2}-\frac{3}{2}}$, which is insignificant for the
undoped case, may play an important role in the doped sample. Its
contribution to the scattering term
to the electron spin coherence is in the similar form
as Eq.\ (\ref{rhoehse}) but all the hole distribution parts are replaced
by terms composed of hh-hh spin coherences. In
the absence of the hole
distribution, it becomes the leading mechanism from the contribution
of EHSE to the spin dephasing. Physically
this contribution is the spin exchange between electrons and virtual
holes. This mechanism, together with other spin dephasing mechanisms
due to the band mixing for the $n$-doped case,\cite{bog} is still
under investigation. A corresponding extension of
the present theory for the $n$-doped material will be published in a
separate paper.

\acknowledgements

MWW would like to thank Prof. L.J. Sham for bringing this topic into
his attention and Prof. A. Imamoglu and Dr. Yutaka Takahashi
for valuable discussions. Dr.
J.M. Kikkawa is acknowledged for providing information about his pertinent
experimental work and critical reading of this manuscript.
This research was supported in part by QUEST, the NSF Science and
Technology Center for Quantized Electronic Structures, Grant  No. DMR
91-20007, and by the National Science Foundation under Grant No. CHE
97-09038 and CDA96-01954, and by Silicon Graphics Inc.

\end{multicols}

\end{document}